\documentclass[aos, preprint, coll]{imsart}
\usepackage[colorlinks,citecolor=blue,urlcolor=blue, bookmarks=false]{hyperref}
\usepackage{bm}
\usepackage{amsthm}
\usepackage{amsmath}
\usepackage{lscape}
\usepackage{amssymb}
\usepackage{tabularx}
\usepackage[dvips]{epsfig}
\usepackage[dvips]{graphics}
\usepackage{graphicx}
\usepackage{dsfont}
\usepackage{url}
\usepackage{slashbox}
\usepackage{natbib}
\usepackage{multirow}
\usepackage{verbatim}
\usepackage[dvipsnames]{xcolor}
\usepackage[normalem]{ulem}
\newtheorem{theorem}{Theorem}[section]
\newtheorem{corollary}[theorem]{Corollary}
\newtheorem{lemma}[theorem]{Lemma}

\theoremstyle{definition}

\makeatletter
\newcommand*{\rom}[1]{\expandafter\@slowromancap\romannumeral #1@}
\makeatother

\newcommand{\Rhat}{\widehat{R}_{j_1, j_2}}

\newcommand{\PP}{\mathbb{P}}

\newcommand{\EE}{\mathbb{E}}
\newcommand{\RR}{\mathbb{R}}

\definecolor{newMaterial}{rgb}{0.5,0.04, 0.175}

\newcommand{\tO}[1]{#1}
\newcommand{\tR}[1]{{#1}}
\newcommand{\ta}[1]{{#1}}

\arxiv{arXiv:1801.07785}

\startlocaldefs
\endlocaldefs

\begin{document}

\begin{frontmatter}

% "Title of the paper"
\title{Strong Sure Screening of Ultra-high Dimensional Data with Interaction Effects\protect}
\runtitle{Strong Screening with Interaction Effects}
%\thankstext[1]{T1}{This work was supported in part by NSF grant DMS-1413366}

\begin{aug}
%%%\title{Strong Sure Screening of Ultra-high Dimensional Categorical Data Exhibiting Trend}
%%%\runtitle{Strong Screening of Categorical Data Exhibiting Trend}

%%%%\begin{comment}
% indicate corresponding author with \corref{}
 \author{\fnms{Randall} \snm{Reese}\ead[label=e1]{randall.reese@inl.gov}}\thanksref{INL,t1},
 \author{\fnms{Xiaotian} \snm{Dai}\ead[label=e3]{xiaotian.dai@aggiemail.usu.edu}}\thanksref{USU}
 \and
 \author{\fnms{Guifang} \snm{Fu}\corref{}\ead[label=e2]{gfu@binghamton.edu}}\thanksref{USU, t2}

 %\author{\fnms{Randall} \snm{Reese}\corref{}\ead[label=e1]{randall.reese@inl.gov}}\thanksref{INL, USU},
%\and
% \author{\fnms{Xiaotian} \snm{Dai}\ead[label=e3]{xiaotian.dai@aggiemail.usu.edu}}\thanksref{USU}

 \affiliation{Idaho National Laboratory\thanksmark{INL} and Binghamton University\thanksmark{USU}}

 %\affiliation

 \thankstext[1]{t2}{Supported by NSF grant DMS-1413366}
 \thankstext[2]{t1}{Corresponding Author}
   
  %%\thankstext[2]{T2}{Corresponding Author}%%Honestly, it is just assumed the first author is the corresponding author. 

\address{Randall Reese\\ Idaho National Laboratory \\ P.O. Box 1625\\Idaho Falls, ID 83415\\
\printead{e1}}

\address{Xiaotian Dai \& Guifang Fu\\Department of Mathematical Sciences\\ Binghamton University\\ P.O. Box 6000\\ Binghamton, NY 13902}%%\\%%Remove } if printing Xiao and GF emails
%%\printead{e3}}
%\phantom{E-mail:\ }\printead*{e3}}

%\affiliation{???}
%\and
%\author{\fnms{???} \snm{???}\ead[label=e2]{???}}
%\address{}
%\affiliation{???}

\runauthor{Reese, Dai, \& Fu}
%%%%%\end{comment}
\end{aug}

%\begin{abstract}
%Ultrahigh dimensional data sets are becoming
%increasingly prevalent in areas such as bioinformatics, medical imaging, and social network analysis. Sure independent screening of such 
%data is commonly used to analyze such data. Nevertheless, few methods exist for screening for interactions among predictors. Moreover, extant 
%interaction screening methods prove to be highly inaccurate when applied to data sets exhibiting strong interactive effects, but weak 
%marginal effects, on the response. We propose a new interaction screening procedure based on joint cumulants which is not inhibited by such 
%limitations. Under a collection of sensible conditions, we demonstrate that our interaction screening procedure has the strong sure screening 
%property. Four simulations are used to investigate the performance of our method relative to two other interaction screening methods. We also 
%apply a two-stage analysis to a real data example by
%first employing our proposed method, and then further examining a subset of selected covariates using multifactor dimensionality reduction.
%\end{abstract}

\begin{abstract}
\ta{The interaction between two features has been recognized to play a pivotal role in contributing to the variation of the response. However, computational feasibility of screening for interactions often is a prohibitive barrier in practice. Extant feature screening approaches are either focused on marginal effects or proven to be highly inaccurate for a scenario presenting strong interactive, but weak marginal, effects. We propose a new interaction screening procedure (referred to as JCIS) based on joint cumulants which is not inhibited by such 
limitations. Four simulations are performed under various conditions to comprehensively demonstrate that JCIS is empirically accurate, theoretically robust, and computationally feasible for ultrahigh dimensional feature spaces. JCIS is also applied to a real data example exhibiting a feature space size yet hereunto unseen in the literature.  We conclude by proving that JCIS possesses the strong sure screening property.}
\end{abstract}



\end{frontmatter}

% AOS,AOAS: If there are supplements please fill:
%\begin{supplement}[id=suppA]
%  \sname{Supplement A}
%  \stitle{Title}
%  \slink[doi]{10.1214/00-AOASXXXXSUPP}
%  \sdatatype{.pdf}"
%  \sdescription{Some text}
%\end{supplement}

%%%%\cite{} to get just year in () yet have the name outside the (), \citep{} to get name and year in ()

\section{Introduction}

%%%%%Start off talking about what we are looking at: INTERACTIONS. 

Ultrahigh dimensional data in fields such as bioinformatics, medical imaging, finance, and the social sciences has become increasingly commonplace. With the yet to cease rapid advances in data collection techniques and computing power, there has arisen an accompanying desire to more comprehensively analyze said data. However, a significant challenge in dealing with ultrahigh dimensional data comes in the fact that classical methods often become intractable or unreliable when confronted with such dimensionality. Here, and throughout this paper, we will use the term \textit{high dimensional} to refer to the case when $p = \mathcal{O}(n^\xi)$ for some constant $\xi>0$ and we will used the term \textit{ultrahigh dimensional} to refer to the case when $\log(p) = \mathcal{O}(n^\xi)$ for some constant $\xi>0$, where $p$ is the number of predictors and $n$ is the number of observations of each of those predictors. 

When the feature space is ultrahigh dimensional, \cite{FanLv:2008} introduce us to the concept of sure independence screening (SIS). Many methods 
possessing the sure screening property of \cite{FanLv:2008} have been developed. Several of these strengthen the original statements of (weak) sure 
screening and establish that the newly proposed method in question has the \emph{strong} sure screening property. See for example \cite{DCSIScitation}, 
\cite{MMLE2010},  and \cite{Huang2014}. However, while there exists (including and beyond those previously mentioned above) an abundance of feature 
screening methods for marginal effects in ultrahigh dimensional feature spaces, \tO{(see for example \citealp{FanFanHiDim:2008,Wang2009forward, 
FanFengSong:2011,ZhuLiLiZhu2011,Kim2012consistent, SuperLongNames,XuChen:2014}, with an overview given in \citealp{Liu2015selective}),} less consideration 
has been given to determining \textit{interactions} between features. 
\ta{Nevertheless, because the relationship between predictors and response is often more complex than can be captured by main effects alone, developing 
techniques for determining interactive effects between predictors is vital. It is common to find two predictors that are simultaneously associated with the 
response and whose effects on the response may not be additive. \tO{In particular, such interaction between variables has been recognized as playing a pivotal 
role in contributing to the variation of the response \citep{HaoZhangiFORM, Fan2015innovated}. As such, procedures addressing interaction screening will allow us to 
better ascertain the interplay between covariates when modelling a response.}}

\ta{One of the most common applications of interaction feature screening for high and 
ultrahigh dimensional data is in the field of bioinformatics and genetics. \citep[See e.g.][]{Phillips2008epistasis, ueki2012ultrahigh, Li2014fast, 
Wei2014detecting, Fang2017tsgsis}. Interactions between single-nucleotide-polymorphisms (SNPs) or interactions between genes are of particular interest. 
(Such genetic interaction is broadly referred to as \emph{epistasis}). 
 With the rapid advances in high-throughput data collection techniques}, detection of epistasis in ultrahigh dimensional data is one very common application of interaction screening.  We will demonstrate a technique for doing such in an empirical analysis given in Section \ref{simulations}. 

\ta{
A significant challenge in the topic of interaction screening, especially when the feature space is high or ultrahigh dimensional, is that of computational feasibility \citep{HaoZhangiFORM}. Many classic approaches designed for detecting interaction may lose power or become increasingly unstable as the number of potential interactions grows \citep[See e.g.][]{Ziegler2007data, Zhang2009willows, Schwarz2010safari, Huang2014, Fan2015innovated}. For a general discussion on the computational challenges of ultrahigh dimensional data, see for example \cite{FanLi:BigData2006} and \cite{FanHanLiu:BigData}.
 Early interaction selection literature such as \cite{Yuan2007efficient}, \cite{Yuan2009structured}, \cite{Zhao2009composite}, and \cite{Choi2010variable} use an approach that is sometimes referred to as \emph{joint analysis}. These methods examine all marginal and interaction effects in a single global search. Such joint analysis approaches are feasible when the feature space is not high dimensional \citep{HaoZhangiFORM}.
However, in many modern interaction screening settings (in which the feature space is likely \emph{not} medium or low dimensional), the most prevalent barrier comes from the growing number of interactions that must be considered.  When the feature space is high or ultrahigh dimensional, and $p$ is also large generally, not only are we faced with the theoretical difficulties of having $p\gg n$, but we also must handle a number of interactions on the order of $\mathcal{O}(p^2)$ \citep{Fang2017tsgsis,Hao2018model}.
 An exhaustive search of all possible interactions requires examining $\binom{p}{2}$ potential interactions for association with the response, something that may not always be feasible to the same degree that the joint analysis approaches were for medium and low dimensional data \citep{HaoZhangiFORM, Fan2015innovated}. 
For even a relatively small value of $p = 10,000$, an exhaustive search requires looking at nearly 50 million potential interactions. When $p$ becomes even larger, the number of interactions to be considered can easily number in the trillions. (For example, $p = 45,000$ would yield over a trillion potential interactions to consider).

Another barrier in the development of interaction feature screening approaches is related to the excessive reliance on an \emph{a priori} heredity 
structure. Specifically, \emph{weak heredity} requires that the interaction between $X_{j_1}$ and $X_{j_2}$ is considered only if at least one of the main 
effects, $X_{j_1}$ or $X_{j_2}$, is individually associated with the response. Similarly, \emph{strong heredity} requires that the interaction between 
$X_{j_1}$ and $X_{j_2}$ is considered only if \emph{both} main effects are influential. Many currently existing approaches rely on determining marginal 
effects first and then examining only a subset of interactions having at least one influential main effect. \citep[See e.g.][]{Li2014fast, HaoZhangiFORM, Hao2018model}. While these approaches significantly reduce the number of interactions that are to 
be examined, they neglect a scenario where there exists an interaction with strong effect, but for which both corresponding main effects are very weak 
\citep{Marchini2005genome}. Handling such instances is of importance in many practical applications \citep{Balding2006tutorial, Marchini2007new, 
Cordell2009detecting, ueki2012ultrahigh, Fang2017tsgsis}.  As such, the ideal interaction screening method will not \emph{a priori} assume that a model must possess a predetermined hereditary structure.}

Herein we will propose an interaction screening method that seeks to address the issues with 
existing interaction screening methods as outlined above. We call this new method JCIS, which stands for ``Joint Cumulant Interaction Screening.''
Unlike many of the aforementioned extant processes for interaction screening, our method will not 
rely on the presence of marginal effects (model heredity) in order to determine interactions between covariates. Moreover, our method is computationally 
more cost effective than current methods, making the application of our proposed techniques to larger and larger data distinctively feasible.  

We will compare JCIS empirically with two existing methods for interaction feature screening: The iterative forward selection method 
(iFORM) of \cite{HaoZhangiFORM} and the generalization of the Pearson Chi-squared-based (PC-SIS) method given by \cite{Huang2014}. 
These two methods most closely resemble our to-be-proposed method in that they admit ultrahigh dimensional data 
and possess certain salient theoretical properties.  

\tO{The iFORM uses an adaptation of forward-selection based on Bayesian information criterion (BIC). This is done by performing a standard stepwise forward selection \citep[see][]{KutnerLinReg2004} and admitting one predictor or interaction at a time. Their approach relies on the 
assumption of strong heredity, as they only admit interactions for which \emph{both} main effects are also present. 
%\cite{Gosik2017iform} presents an 
%implementation of iFORM in the setting of epistasis.% Nice but irrelevant in our case. 

The mathematical generalization of the PC-SIS method given by \cite{Huang2014} will be referred to as Generalized Pearson Correlation 
(GPC). While GPC does not directly require selecting marginal effects before proceeding with screening for interactive effects, their method can become 
computationally intractable (by their own admission) when applied to any exhaustive search of all possible interactions. (Under empirical observation, our newly proposed method ran about six to seven times faster than GPC when applied to the same data sets on the same machine). As such, GPC has never been 
numerically tested under a setting where marginal features have not first been selected. The empirical results of Section \ref{simulations} will 
furthermore show that GPC produces rather unfavorable results when main effects are not first determined. 
%Our newly proposed method overcomes many of the 
%less than desirable issues of both iFORM and GPC.
}

The remainder of this paper is outlined as follows: In Section \ref{theory}, we will detail our proposed interaction screening method, including statements on the theoretical properties for said method.
This is followed by several numeric simulations in Section \ref{simulations}. Among many applications of interaction screening, perhaps the most prevalent is in genome-wide association studies (GWAS) for determining interactive (epistatic) effects between single nucleotide polymorphisms (SNPs). We too will demonstrate such an application in a real data analysis, also found in Section \ref{simulations}. In Section \ref{discuss} we provide concluding remarks on our method and our findings. Finally, Section \ref{proofSection} contains the proofs of the theorems presented in Section \ref{theory}.

\section{Interaction Screening using Joint Cumulants}\label{theory}

\subsection{Some Preliminaries}

\ta{Let $(Z_1, Z_2, \ldots, Z_r)$ be a $r$-dimensional random vector. The $r$-way joint cumulant of $(Z_1, Z_2, \ldots, Z_r)$ is defined as 
\[\kappa_r(Z_1, Z_2, \ldots, Z_r) = \EE\left[\prod_{i = 1}^r(Z_i - \EE Z_i)\right]. \]
Defined this way, the joint cumlant is an $r$-way analogue of the covariance. (In fact, the 2-way joint cumulant is the standard covariance function). }

\ta{Specific to our purposes here, the three-way joint cumulant, written as $\kappa_3(\cdot, \cdot, \cdot)$, is given as follows:
\begin{eqnarray*}\kappa_3(Z_1, Z_2, Z_3) &=& \EE\left[(Z_1 - \EE Z_1)(Z_2 - \EE Z_2)(Z_3 - \EE Z_3)\right]\\
&=& \EE(Z_1Z_2Z_3) - \EE(Z_1Z_2)\EE Z_3-\EE(Z_1Z_3)\EE Z_2\\
&& \quad \quad \quad \quad \quad ~-\EE(Z_2Z_3)\EE Z_1 + 2 \EE Z_1\EE Z_2\EE Z_3.\end{eqnarray*}}Note that $\kappa_3(Z_1, Z_2, Z_3)$ is zero if any one of the variables is statistically independent from the other two.

\ta{\subsection{An Interaction Ranking and Screening Procedure}
In this section we propose a new interaction screening procedure. Let $Y$ be the response variable and $X_j (j=1,\ldots, p)$ be the $j^{th}$ predictor, with $p$ being the total number of predictors. 
}

\ta{Inspired by the Pearson correlation of two random variables, we propose a new generalized form, $R_{j_1, j_2}$ $(j_1<j_2;~j_1,j_2=1,\ldots,p)$}, which admits \emph{three} random variables as arguments:
\begin{equation}
R_{j_1, j_2} = \frac{|\kappa_3(Y, X_{j_1}, X_{j_2})|}{\sqrt{\kappa_2(X_{j_1},X_{j_1} )\kappa_2(X_{j_2}, X_{j_2})\kappa_2(Y,Y)}}.
\label{Rjj}
\end{equation}
We estimate $R_{j_1, j_2}$ as
\begin{equation}
\hat{R}_{j_1, j_2}=\frac{\sqrt{n}\left|\sum\limits_{i = 1}^n(X_{ij_1} - \overline{X}_{j_1})(X_{ij_2} - \overline{X}_{j_2})(Y_i - \overline{Y})\right|}{\sqrt{\left(\sum\limits_{i = 1}^{n}(X_{ij_1} - \overline{X}_{j_1})^2\right)\left(\sum\limits_{i = 1}^{n}(X_{ij_2} - \overline{X}_{j_2})^2\right)\left(\sum\limits_{i = 1}^n(Y_i-\overline{Y})^2\right) }},
\label{Rhat}
\end{equation}
\ta{where $X_{ij_1}$, $X_{ij_2}$, and $Y_i$ are sample observations and $\overline{X}_{j_1}$, $\overline{X}_{j_2}$, and $\overline{Y}$ are sample means of $X_{j_1}$, $X_{j_2}$, and $Y$, respectively.}

\ta{Let $\mathcal{S}_F = \{(j_1,j_2)~|~j_1< j_2;~j_1,j_2=1,\ldots,p\}$ denote the indices of the full model, which refers to all pairs predictors except the \tR{given} predictor with itself. Let $\mathcal{S} \subseteq \mathcal{S}_F$ denote the indices of an arbitrary interaction model under consideration. \tR{Designate by $\bm{X}_{(\mathcal{S})}$ the interaction model containing all two-way interactions between the covariate pairs in $\mathcal{S}$.} When we need to refer to the interaction between $X_{j_1}$ and $X_{j_2}$, we will use the notation $(X_{j_1}, X_{j_2})$. }

Given some model $\mathcal{S}$, we will let $\mathcal{D}\left(Y_i \mid \bm{X}_{\mathcal{S}}\right)$ indicate the conditional distribution of $Y_i$ given the covariates of $\bm{X}_{\mathcal{S}}$. A model $\mathcal{S}$ will be considered sufficient if
\[\mathcal{D}\left(Y_i \mid \bm{X}_{\mathcal{S}_F}\right) = \mathcal{D}\left(Y_i \mid\bm{X}_{\mathcal{S}}\right)\]
The full model $\mathcal{S}_F$ is of course trivially sufficient. We are ultimately interested only in finding the sufficient model with the fewest number of interaction pairs. We will call the smallest sufficient model (i.e. the sufficient model with the least number of pairs) the true model. Our aim overall is to determine an estimated model which contains the true model and is moreover the \textit{smallest} such model to contain the true interaction features.
The following section will outline the specifics of our proposed interaction screening approach for estimating the true model. As a matter of further notation, we will denote the true model by $\mathcal{S}_T$ and an estimated model by $\widehat{\mathcal{S}}$.

We form the estimated model $\widehat{\mathcal{S}}$ by choosing some cutoff $c >0$. Methods for choosing such a $c$ are varied (see for example \cite{FuWangDai2017}) and will not be the focus of this paper. Define $\widehat{\mathcal{S}}$ as follows:
\[\widehat{\mathcal{S}} = \{j: 1\leq j \leq p, ~ \Rhat > c\}.\]
Designate the numerator of $\Rhat$ (sans the constant $\sqrt{n}$ and the absolute values) by $\hat{\tau}_{j_1, j_2}$. We will show that $\hat{\tau}_{j_1, j_2}$ is a consistent estimator of the numerator of $R_{j_1, j_2}$, $\kappa_{3}(X_{j_1}, X_{j_2}, Y)$. The denominator of $\Rhat$ consists of (biased) sample estimators for the standard deviations of $X_{j_1}$, $X_{j_2}$, and $Y$. (The bias of these estimators will disappear asymptotically, however). It is a routine proof to show that these estimators of the standard deviations are consistent estimators of their respective standard deviations.

\subsection{Theoretical properties}\label{thrtProp}
We establish four conditions that will aid us in determining further properties of JCIS:
\begin{enumerate}
    \item[(C1)] \emph{Lower bound on the standard deviations}. We assume that there exists a positive constant $\sigma_{\text{min}}$ such that for all $j$,\[\sigma_j > \sigma_{\text{min}}\quad\quad\text{and}\quad\quad\sigma_Y > \sigma_{\text{min}}\]
    This excludes features that are constant and hence have a standard deviation of 0.
   \item[(C2)] \emph{Upper bound on the standard deviations}. We take as our second condition the assumption that \[\sigma_j, \sigma_Y < \sigma_{\text{max}} < \infty\] for all $j= 1,2,3,\ldots, p$. This is a relatively lenient condition, and one that is easy to satisfy in a large variety of applications. When each of $X_{j_1}$, $X_{j_2}$, and $Y$ are categorical  and ordinal (with $Y$ being binary and each covariate having, without loss of generality, $K$ many levels), we can explicitly obtain a simultaneous upper bound on each $\sigma_{j_1}, \sigma_{j_2}, $ and $\sigma_Y$ by use of Popoviciu's inequality on variances (see \cite{Popoviciu}):
\[\text{Let } \quad \sigma_{\text{max}} = \text{max}\left\{\frac{1}{2},~ \sqrt{\frac{1}{4}\left(v_{K}-{v}_{1}\right)}\right\},\]
where the first term in the maximum selection is a bound on the standard deviation of $Y$ and the second term is given by Popoviciu's inequality on variances. Here $v_1$ and $v_K$ represent the lowest and the highest levels (by chosen encoding) of any $X_j$.
    %%%The citation for Popoviciu's inequality on variances is

%%%%%%Popoviciu, T. (1935). "Sur les équations algébriques ayant toutes leurs racines réelles". \textit{Mathematica (Cluj)} 9: 129–145.

    \item[(C3)] \textit{Joint cumulant association}.     Define the following function on a subset of $\RR^3$:
    \[\omega_{j_1, j_2}(k_1, k_2, m) = |(k_1 - \EE X_{j_1})(k_2 - \EE X_{j_2})(m - \EE Y)\pi_{j_1, j_2, Y}(k_1, k_2, m)|,\]
    where $\pi_{j_1, j_2, Y}$ is the joint probability density function for $X_{j_1}$, $X_{j_2}$, and $Y$. Assume that $\omega_{j_1, j_2}(k_1, k_2, m)$ is of the same sign for all $(k_1, k_2, m)$. Without loss of generality, we will assume a positive sign in each instance. 
    Taking $X_{j_1}$ and $X_{j_2}$ as having the same support $\Psi \subseteq \RR$, we now assume there exists a positive constant $\omega_{\text{min}}$ such that \[\min_{(j_1, j_2) \in \mathcal{S}_T}\left(\sup_{\substack{k_1, k_2 \in \Psi \\m \in \RR}}\left\{\omega_{j_1, j_2}(k_1, k_2, m)\right\}\right) > \omega_{\min}>0.\]
This is an easy assumption to require the true features to satisfy and should be quite easy to achieve in a wide variety of reasonable situations.

    \item[(C4)] \textit{Existence}. \tO{Assume that $R_{j_1, j_2} = 0$ for any pair of indicies $(j_1, j_2) \not\in \mathcal{S}_T$.} It is also to be assumed that $R_{j_1, j_2}$ exists for all $X_{j_1}$ and $X_{j_2}$ pairs. That is, $R_{j_1, j_2} < \infty$. 
\end{enumerate}

We can now state the following theorems:

\begin{theorem}\label{Thm1} (\textit{Strong Screening Consistency}). Given conditions (C1), (C2), (C3) and (C4), there exists a positive constant $c>0$ such that \[\mathbb{P}(\widehat{\mathcal{S}} = \mathcal{S}_T) \longrightarrow 1 \text{ as } n \longrightarrow \infty.\]
\end{theorem}
\begin{theorem}\label{Thm2} (\textit{Weak Screening Consistency}). Given  that conditions (C1), (C2), and (C3) still hold, while removing from (C4) only the assumption of $R_{j_1, j_2} = 0$ for all $(j_1, j_2) \notin \mathcal{S}_T$, there exists a positive constant $c>0$ such that \[\mathbb{P}(\widehat{\mathcal{S}} \supseteq \mathcal{S}_T) \longrightarrow 1 \text{ as } n \longrightarrow \infty.\]
(But $\mathbb{P}(\widehat{\mathcal{S}} \subseteq \mathcal{S}_T)$ may not converge to 1 as $n$ approaches infinity).
\end{theorem}
The proofs of these two theorems are presented in Section \ref{proofSection}.

%%%%%%%%%%%%%%%%%%%%%%%%%%%%%%%%%%%%%%%%%%%%%%%%%%%%
%%%% COROLLARIES
%%%%%%%%%%%%%%%%%%%%%%%%%%%%%%%%%%%%%%%%%%%%%%%%%%%%
\subsection{Corollaries}
We can draw several corollaries from the proofs of Theorems \ref{Thm1} and \ref{Thm2} (see Section \ref{proofSection}). These results do not themselves directly deal with sure screening, but they nevertheless allow us to make observations pertaining to the underlying mechanics of JCIS.

\begin{corollary}\label{Cor1}%%%Make this a theorem?
In the initial step of the proofs of Theorems \ref{Thm1} and \ref{Thm2}, it will be shown that there exists a value $R_{\min}$ such that for any pair $(j_1, j_2) \in \mathcal{S}_T$, we have $R_{j_1, j_2} > R_{\min}$.
\end{corollary}

\begin{corollary}\label{Cor2}
From the end of Step 2 in the proofs of Theorems \ref{Thm1} and \ref{Thm2}, we will conclude that $\Rhat$ converges \textit{uniformly} in probability  to $R_{j_1, j_2}$. In other words,
\[\mathbb{P}\left(\max_{(j_1, j_2)}|\Rhat - R_{j_1, j_2}| > \varepsilon\right) \rightarrow 0 \quad \text{ as } n\rightarrow\infty\] for any $\varepsilon> 0$.
\end{corollary}

\tO{
\section{Simulations and Empirical Data Analysis}\label{simulations}
We compared our method to iFORM \citep{HaoZhangiFORM}, as well as to the generalized PC-SIS method (which they leave unnamed, but which we have call GPC) of \cite{Huang2014}.
\tR{Which method we compared JCIS to depended on the data type of the response. GPC admits only categorical responses; iFORM admits only continuous responses. JCIS allows for either categorical or continuous responses, which in and of itself is salient. In Simulations 1 and 2, we report the median and mean ranking of all causative interactions, as found by both JCIS and GPC. The closer these scores were to one, the more accurate the method can be said to be. The first of these simulations shows the prodigious failure of GPC to detect interactions when faced with a true model containing no marginal effects. The second of these simulations demonstrates that even when marginal effects may be present, if they are not first determined, GPC will yet fail to determine the causative interactions with any degree of accuracy. 
In Simulations 3 and 4, we report the percentage of replicates for which a given method found the causative interactions to be among the top five most important interactions. The closer to 100\% the percentage was, the more accurate the method can be said to be. As was the case with GPC, iFORM struggled in both these simulations to accurately determine the causative interactions when the marginal heredity structure is unknown.
Each of these simulations, as well as the associated results, are summarized below. %%%(See Section \ref{simTables} for the results).
 \tO{We also performed an analysis on an empirical data set relating to polycystic ovary syndrome (PCOS) from the NCBI databases.}}

\subsection{Simulation 1}
In this simulation, we will be observing $200$ samples ($n = 200$) of $1000$ covaraiates ($p = 1000$). Of these $p$-many covariates, only the interaction between $X_1$ and $X_2$ will be considered to have meaningful contribution to the outcome $Y$. We will run 100 replications and report the average (mean) ranking and the median ranking of the interaction between $X_1$ and $X_2$ as it relates to association with $Y$. The test data is the same for both JCIS and GPC.

We generate all $X_j$ randomly from the set $\{0,1\}$, with each outcome being equally likely. We then let \[Y = X_1 \times X_2.\] This will mean that $Y$ depends on only the interaction between $X_1$ and $X_2$ (i.e. $\mathcal{S}_T = \{(1,2)\}$). Note that we omit any main effects to exhibit the robustness of JCIS even in the absence of main effects on the response.
The results for Simulation 1 are summarized in Table \ref{table:Sim1} below.

\begin{table}[h!]
\caption{Mean and Median Ranking of Interaction Between $X_1$ and $X_2$ in Simulation 1}\label{table:Sim1}
\begin{center}
\begin{tabular}{ |c | c c | }
\hline
 & JCIS & GPC  \\ \hline\hline
Mean Rank of $(X_1, X_2)$ & 1 & 2104.5 \\
Median Rank of $(X_1, X_2)$ &1 &1306\\
  \hline
\end{tabular}
\end{center}
\end{table}
Note that GPC fails prodigiously to establish the importance of the interaction between $X_1$ and $X_2$ on the response $Y$. Both the average and the median rankings of $(X_1, X_2)$ by GPC are much too large for GPC to be considered a reliable feature screening approach in this case. On the other hand, our JCIS method accurately ranks $(X_1, X_2)$ as being the most important interaction in relation to the response in each of the 100 replicates.

}

\tO{
\subsection{Simulation 2} This simulation closely resembles the interaction simulation found in \cite{Huang2014}. Here, we assume that $Y$ only has two levels. (The original simulation assumes $Y$ has four levels). We also retain the assumption in the aforementioned simulation of \cite{Huang2014} that each $X_j$ is binary. It should be noted that, while GPC performs admirably in the original simulation of \cite{Huang2014}, that simulation is apparently dependent on first identifying a small set of relevant main effects, something that we do not do here. This demonstrates one marked benefit of JCIS over GPC: No predetermined set of predictors is required to obtain accurate results. This holds true whether important causative main effects exist or not. \tO{To be specifically clear, GPC does not explicitly require obtaining main effects first. This means that their method \emph{in theory} works for cases where the marginal effects are weak and the interactive effects are strong. However, the results of this simulation here will show that the GPC algorithm is wildly inaccurate in this case.} 

We will be observing $200$ samples ($n = 200$) of $1000$ covariates ($p = 1000$). First, we generate a response vector $Y$, where $Y = 0$ or $Y =1$ and $\PP(Y_i = 1) = 0.75$. Next, we generate $X_{ij} \in \{0,1\}$ for $j = 1,3,5,7$ as follows:
\begin{itemize}
\item[$\bullet$] Conditional on $Y_i = k$, let $\PP(X_{ij} = 1 | Y_i = k) = \theta_{kj}$, where $\theta_{kj}$ is given in Table \ref{table:Sim2Theta}.

\begin{table}[h]
\caption{$\theta_{kj}$ Values for Simulation 2}\label{table:Sim2Theta}
\centering
\begin{tabular}{|l| cccc|}
\hline
 & \multicolumn{4}{c|}{j} \\ \cline{2-5}
$\theta_{kj}$ & 1 & 3 & 5 & 7 \\ \hline\hline
$k = 0$ & 0.3 & 0.4 & 0.5 & 0.3 \\
$k = 1$ & 0.95 & 0.9 & 0.9 & 0.95 \\ \hline
\end{tabular}
\end{table}

\item[$\bullet$] Given $Y_i$ and $X_{i, 2m-1}$ (for $m = 1,2,3,4$), we generate $X_{i,2m}\in \{0,1\}$ using the following probabilities:
\small
\[\PP(X_{i, 2m} = 1 | Y_i =k, X_{i, 2m-1} = 0) = 0.6I(\theta_{k,2m-1} > 0.5) + 0.4I(\theta_{k,2m-1} \leq 0.5);\]
\[\PP(X_{i, 2m} = 1 | Y_i =k, X_{i, 2m-1} = 1) = 0.95I(\theta_{k,2m-1} > 0.5) + 0.05I(\theta_{k,2m-1} \leq 0.5),\]
\normalsize where $I(\cdot)$ is the standard indicator function.

\item[$\bullet$] For all remaining covariates (i.e. $X_j$ for $j > 8$), randomly sample the set $\{0,1\}$ with $\theta_{kj} = 0.5$.
\end{itemize}
Overall, the causative interactive effects will be $(X_1, X_2)$, $(X_3, X_4)$, $(X_5, X_6)$, and $(X_7, X_8)$. The test data is the same for both JCIS and GPC. We will run 100 replications and report the average (mean) ranking and the median ranking of each these interactions as they relate to association with $Y$.
The results of Simulation 2 are given in \ref{table:Sim2} below.

\begin{table}[h!]
\caption{Mean and Median Ranking of Causative Interactions in Simulation 2}\label{table:Sim2}
\begin{center}
\begin{tabular}{ |c | c c | }
\hline	
 & JCIS & GPC  \\ \hline\hline
Mean Rank of $(X_1, X_2)$ & 2.01 & 7302.73 \\
Median Rank of $(X_1, X_2)$ &2 &534\\\hline
Mean Rank of $(X_3, X_4)$ & 3.53 & 2365.05 \\
Median Rank of $(X_3, X_4)$ &3 &42.5\\\hline
Mean Rank of $(X_5, X_6)$ & 4.65 &936.65 \\
Median Rank of $(X_5, X_6)$ &4 &16.5\\\hline
Mean Rank of $(X_7, X_8)$ & 2.33& 6563.83 \\
Median Rank of $(X_7, X_8)$ &2 &1083.5\\
  \hline
\end{tabular}
\end{center}
\end{table}

Since it is obviously impossible for every causative interaction to be consistently ranked as the absolute top interaction, any method placing each true interaction on average in the top four or so causative interactions can easily be said to be producing accurate results. In this regard, JCIS obtains excellent results.
However, as has been mentioned previously, we also see the unfortunate over-reliance of GPC on first establishing a small set of relevant main effects in order to produce a dependable set of causative interactions. The average ranking of each causative interaction by GPC does not lend to confidence in being able to select via GPC the true interactions with any degree of consistency. Although the median rank of each interaction by GPC is better (and even decent in the case of $(X_5, X_6)$) than the average respective rank by GPC, the reliability and stability of the method is, on the whole, questionable. The large difference between the mean and median rankings associated with GPC indicates that GPC was wildly unstable over the course of 100 replications. Combined with the aforementioned inabilities of GPC to accurately rank the interactions, JCIS proves to be the superior method under these settings.   

}

\tO{
\subsection{Simulation 3}
Simulation 3 is similar in form to Simulation 1. However, we now will test the ability of JCIS to screen for interactions when the covariates are continuous.
We will be observing $200$ samples ($n = 200$) of $1000$ covariates ($p = 1000$). Of these $p$-many covariates, only the interaction between $X_1$ and $X_2$ and the interaction between $X_3$ and $X_4$ will be considered to have meaningful contribution to the outcome $Y$.

We generate all $X_j$ randomly based on repeated random samples of the normal distribution with mean 0 and standard deviation 2:
\[X_j \overset{\text{i.i.d.}}{\sim} N(\mu = 0, \sigma = 2).\] We then let \[Y = X_1 \times X_2 + X_3 \times X_4.\] Note that this simulation will also be testing the ability of JCIS to correctly locate multiple two-way interactions having an effect on the response. We will report the percentage of replicates (out of 100) where the interactions $(X_1, X_2)$ and $(X_3, X_4)$ are individually within the top five interactions detected, as well as the percentage of time that \emph{both} $(X_1, X_2)$ and $(X_3, X_4)$ are simultaneously within the top five interactions detected.
The test data is the same for both JCIS and the iFORM method of \cite{HaoZhangiFORM}. The results of Simulation 3 are detailed in Table \ref{table:Sim3}.

\begin{table}[h!]
\caption{Percentage of Replicates Finding $(X_{j_1}, X_{j_2})$ to be Important in Simulation 3}\label{table:Sim3}
\begin{center}
\begin{tabular}{ |c | c c | }
\hline
 & JCIS & iFORM  \\ \hline\hline
$(X_1, X_2)$ &100\% &0\%  \\
$(X_3, X_4)$ &100\% & 0\% \\
$(X_1, X_2)$ \& $(X_3, X_4)$ & 100\% & 0\%\\
  \hline
\end{tabular}
\end{center}
\end{table}

Here an interaction is considered to be ``important'' if it is ranked in the top five most relevant interactions by the screening method in question.
These results show the remarkable difference between JCIS and iFORM in being able to determine interactive effects in the event that no main effects are prevalent in the data. This demonstrates the previously discussed limitation of iFORM in requiring the existence of main effects between covariates in order to find any meaningful interactive effects. This is especially important when one wants to screen for interactions in genetic data, where gene SNPs with weak marginal effects can have stronger interactive effects on the response. For further discussion on this, see e.g. \cite{manolio2009} and \cite{ueki2012ultrahigh}. 

}

\tO{
\subsection{Simulation 4}
In this simulation we will test the ability of JCIS versus iFORM in successfully screening two interactive features in the presence of individual main effects among those covariates forming the interactive effects. This is done in order to show that even when strong marginal effects are present, JCIS can outperform iFORM in determining the true interactions. We examine 100 observations ($n = 100$) of 500 covariates ($p = 500$) over 100 replications.
Let $X$ follow the multivariate normal distribution with mean vector \textbf{0} and $\text{cov}(X_{j_1}, X_{j_2}) = 0.1^{|j_1 - j_2|}$ for $1 \leq j_1, j_2 \leq p$. Now define \[Y = X_1 + X_3 + X_6 + X_{10} + 3(X_1 \times X_3) + 3(X_6 \times X_{10}).\]
We will apply JCIS and iFORM to screen for the true interactions $(X_1, X_3)$ and $(X_6, X_{10})$.
 The results of Simulation 4 are given in Table \ref{table:Sim4}.

 \begin{table}[h!]
\caption{Percentage of Replicates Finding $(X_{j_1}, X_{j_2})$ to be Important in Simulation 4}\label{table:Sim4}
\begin{center}
\begin{tabular}{ |c | c c | }
\hline
 & JCIS & iFORM  \\ \hline\hline
$(X_1, X_3)$ &92\% &21\%  \\
$(X_6, X_{10})$ &92\% & 19\% \\
$(X_1, X_3)$ \& $(X_6, X_{10})$ & 84\% & 9\%\\
  \hline
\end{tabular}
\end{center}
\end{table}

Here again, an interaction is considered to be ``important'' if it is ranked in the top five most relevant interactions by the screening method in question.
Note that even when marginal effects are included in the generation of the response, iFORM still struggles to accurately and consistently detect the true interactive effects. Note that JCIS, on the other hand, accurately detects at least one of the two true interactions in every replication, and detects \emph{both} true interactions in 84 of the 100 replications. Overall, this simulation suggests that even when marginal effects are present (a scenario which is more favorable for iFORM), JCIS still is the superior method. 
}

\subsection{Final Comments on Simulation Results}
An overall issue that arises in interaction feature screening is the reliance of extant methods on a predetermined set of marginally important predictors. Simulations 1 and 3 demonstrate this shortcoming in even very simple cases. The first and third simulations lead us to believe that, in the absence of strong main effects, JCIS is a much superior method to GPC and iFORM.
Simulations 2 and 4 add main effects to the simulation data model. However, even with the presence of main effects, both GPC and iFORM do not produce competent or reliable results. Again, as with the first and third simulations, JCIS performs admirably.

\subsection{Real Data Analysis: Epistasis Detection}
We apply a two-stage process to a real data set examining prevalence of polycystic ovary syndrome (PCOS) in females who self-identified as having Caucasian or European-ancestry. With the proper approvals, this PCOS dataset was downloaded from the database of genotypes and phenotypes (dbGaP) of the National Center for Biotechnology Information (NCBI) at the NIH (dbGaP Study Accession: \href{https://www.ncbi.nlm.nih.gov/projects/gap/cgi-bin/study.cgi?study_id=phs000368.v1.p1}{phs000368.v1.p1}). This data consists of 4099 (3055 controls, 1042 cases) observations of each of 731,442 SNPs. The response is PCOS affection status (0 = control, 1 = case) and the predictors are the encoded SNP geneotype values (where 1, 2, and 3 correspond with homozygous recessive, heterozygous, and homozygous dominant respectively).  Our specific aim is to identify SNPs which most strongly interact with one another in determining PCOS affection status.

\subsubsection{Stage 1 analysis} In the first stage of our interaction feature screening, we apply JCIS to all pairwise combinations of SNPs coming from the same chromosome. All 23 \textit{homo sapien} chromosomes were used. As is common with such data sets, we removed all SNPs with less than a 95\% call rate, as well as all SNPs with a minor allele frequency less than 10\%. \citep[See][]{anderson2010data, li2010bayesian, btq272}. All of the analysis in this stage was performed on the cluster machine centered at the Center for High Performance Computing at the University of Utah. After recording a $\widehat{R}_{j_1, j_2}$ value for all possible within-chromosome pairs, we ordered the SNP pairs from largest to smallest
$\widehat{R}_{j_1, j_2}$ value. To ensure that all important SNP pairs are selected after the first stage, we keep all SNPS associated with the $n = 4099$ largest values of $\widehat{R}_{j_1, j_2}$. We then proceed to Stage 2 of the real data analysis.

It should here be noted that while an exhaustive search among \textit{all} pairs of SNPs (including between-chromosome pairings) can be done, preliminary results on all possible between-chromosome pairs of SNPs from chromosomes 11 through 23 indicated that approximately 300,000 within-chromosome SNP pairings (SNP pairs coming from the same chromosome) had a stronger interactive effect on PCOS status than even the top ranked between-chromosome SNP pair. Further examination as to why this is could be pursued at a later date.

\subsubsection{Stage 2 analysis}
We now turn our attention to a more in depth analysis using multifactor dimensionality reduction (MDR) on a small set of the SNPs comprising SNP pairs having the largest values of $\Rhat$. MDR is a model-free and nonparametric approach first introduced in \cite{RITCHIEmdr2001} that can be used to identify high-order SNP-SNP interactions, even in the absence of independent main effects of the gene SNPs on the outcome. Ideally, we would like to select a set of SNPs associated with the $n$ largest $\Rhat$ values. However, as discussed in much of the MDR literature (e.g. \cite{RITCHIEmdr2001}, \cite{Hahnmdr2003}, \cite{Winham2011}, \cite{GolaRoadmapMDR2016}), the run time of MDR increases drastically as the number of SNPs under consideration grows. Thus, we must choose a relatively small set of SNPs to consider for analysis by MDR. Therefore, we keep only the top 85 SNP pairs having the largest $\Rhat$ values. This corresponds with a cutoff value of $\Rhat > 0.865$, at which there is a distinct clustering of $\Rhat$ values. 

%In order to select a reasonably small set of SNPs to consider (approximately 100 candidate SNPs), we can project onto the position of $\text{SNP}_{j_1}$ 
%the top 10,000 $\Rhat$ values, then look for a cutoff value for which most of the SNPs lie below. We only plot the top 10,000 $\Rhat$ values due to 
%computational limitations in plotting a larger set of values. Approximately 99\% of the $\Rhat$ values are less than $0.1$. This tells us that the vast 
%majority of SNP pairs can be omitted as having little to no effect on PCOS status. The top 10,000 SNPs  still easily provide a set of SNPs encompassing the %overall patterns of the $\Rhat$ values.

% Figure \ref{fig:RhatVals} shows the top 10,000 $\Rhat$ values versus the associated position of  $\text{SNP}_{j_1}$. We will look for a cutoff value for which (approximately) less than 100 $\Rhat$ values lie above.
%\begin{figure}[h]
%\centering
%    \includegraphics[scale=0.64]{RhatVSsnp_1}
%    \caption{$\text{SNP}_{j_1}$ position versus $\Rhat$ value.}
%    \label{fig:RhatVals}
%\end{figure}
%Based on the plot in Figure \ref{fig:RhatVals}, we select a cutoff of $\Rhat = 0.865.$ This yields a computationally feasible set of 85 SNPs for our candidate set for MDR.

Because our case to control ratio is unbalance (i.e. not equal to 1), we will use balanced accuracy (BA) as the evaluation measure of our MDR results. The BA can be defined as follows:
\small
\[BA = \frac{1}{2}\left(\frac{\text{True Positives}}{\text{True Positives} + \text{False Negatives}} +\frac{\text{True Negatives}}{\text{True Negatives} + \text{False Positives}}\right), \]
\normalsize where the true and false positives and negatives refer to the classification of a subject based on the loci-genotype combinations selected by MDR. Note that BA is just the arithmetic mean of the specificity and the sensitivity. For further discussion on the use of the BA as the metric for our model evaluations in the presence of unbalanced case to control ratios see, for example, \cite{MDRunbalanced} and \cite{Winham2011}.

A further consideration that must be made in regards to the unbalanced ratio of case to control PCOS instances is that of choosing a threshold $T$ at which to classify subjects as high or low risk for PCOS based on their genotype combination among the SNPs selected by MDR. While traditional approaches tacitly assume \emph{a priori} balance of the case to control ratio (either by design or by over/undersampling of the case/control observations), and take $T = 1$ as the threshold, more robust implementations of MDR allow for an adjusted threshold $T_{adj}$, where $T_{adj}$ is the case to control ratio. This use of the adjusted threshold is seen commonly in the MDR literature (e.g. \cite{MDRunbalanced} and \cite{GolaRoadmapMDR2016}).

Using an implementation of MDR in Java from the researchers at www.epistasis.org (see also \cite{PattinMDR2009} and \cite{GreeneMDR2010}, both of which recommend this implementation), we obtained the following two, three, four, and five-loci results. Table \ref{table:MDR} contains the 10-fold cross validated accuracies (BA-wise) for each model. 
\begin{itemize} %% 
\item[$\bullet$] Two-loci model:\texttt{rs1024216}, \texttt{rs1423304}. 

\item[$\bullet$] Three-loci model: \texttt{rs1002424}, \texttt{rs1423304}, \texttt{rs1024216}.

\item[$\bullet$] Four-loci model: \texttt{rs1002424, rs1024216, rs657718, rs4745466}. 

\item[$\bullet$] Five-loci model: \texttt{rs1002424, rs1423304, rs1024216, rs657718, rs4745466}.
\end{itemize}%

\begin{table}[h!]
\caption{MDR Accuracy}\label{table:MDR}
\begin{center}
\begin{tabular}{ |c |  c | }
\hline
k-way & CV Accuracy\\ \hline\hline
2-way & 53.50\%\\
3-way & 52.69\%\\
4-way & 51.97\%\\
5-way &53.26\%\\
  \hline  
\end{tabular}
\end{center}
\end{table}

\begin{figure}[h]
\centering
    \includegraphics[scale=1]{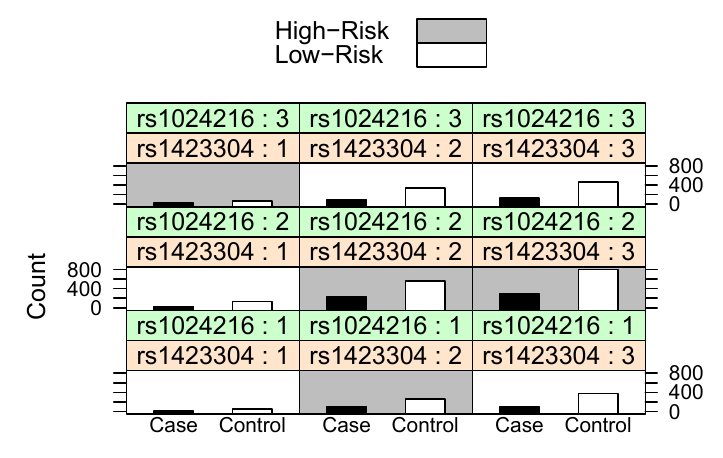}
    \caption{High-low risk bar plots broken down by genotype for the selected two-loci model. A total of 10 subjects were removed from consideration due to missing genotype information at the associated SNPs.}
    \label{fig:2wayMDR}
\end{figure}

\begin{landscape}
\begin{figure}
\centering
    \includegraphics[scale=0.55]{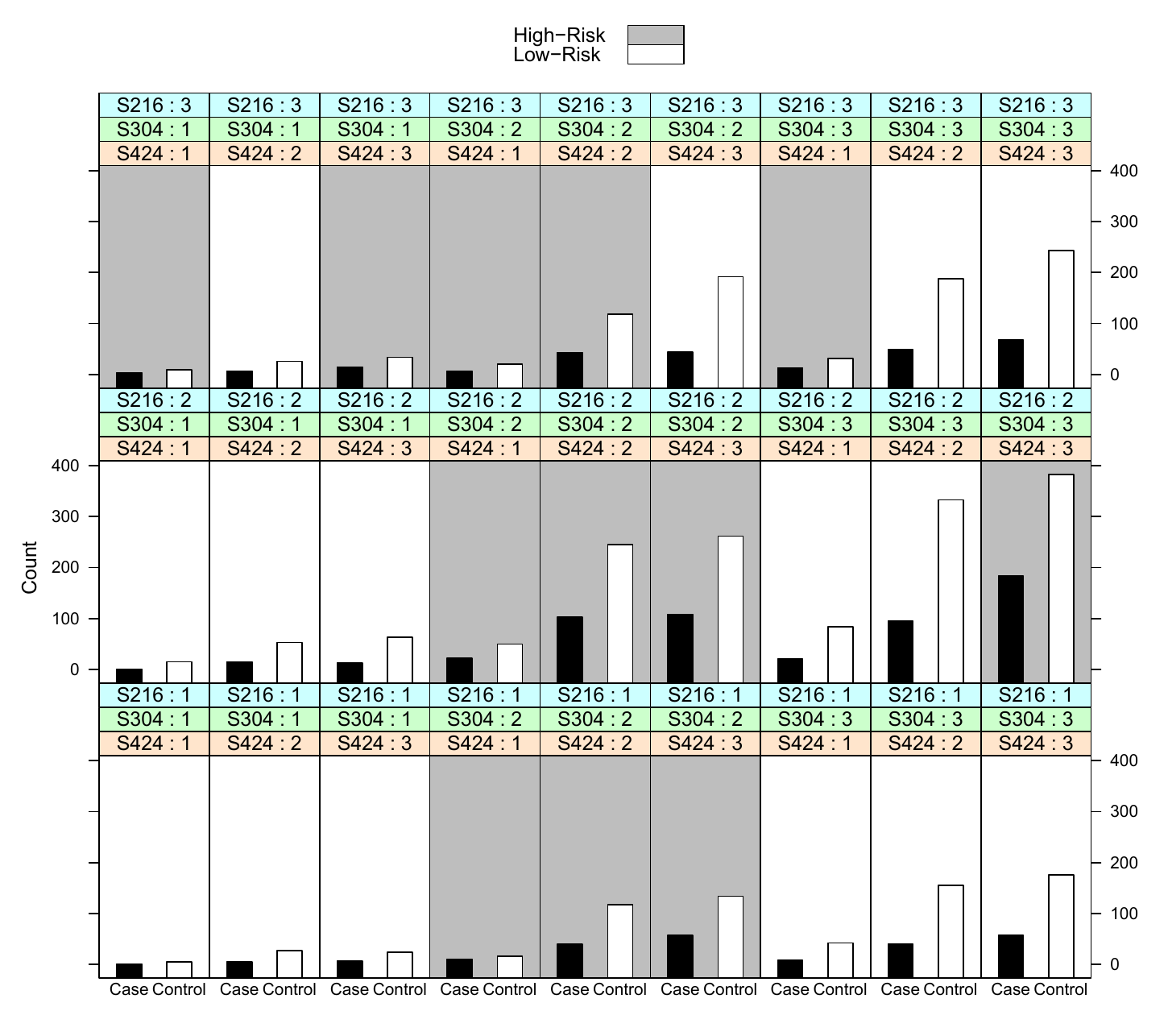}
    \caption{High-low risk bar plots broken down by genotype for the selected three-loci model. Here ``S---'' indicates the selected SNP ending in the given three digits. A total of 17 subjects were removed from consideration due to missing genotype information at the associated SNPs.}
    \label{fig:3wayMDR}
\end{figure}
\end{landscape}

All models found via this MDR implementation employ the ensemble of BA to test model accuracy, 10-fold cross validation to prevent overfitting, and the adjusted threshold outlined by \cite{MDRunbalanced}. Higher order models can also be found; however, with greater balanced accuracy comes an exponentially increasing computational cost. The plots given in Figures \ref{fig:2wayMDR} and \ref{fig:3wayMDR} show the classification of geneotype combinations in the two- and three-loci models as either high or low risk for PCOS. Genotype combinations associated with low-risk for PCOS are indicated with a white background on the MDR bar plot; genotype combinations found to be associated with high-risk for PCOS are denoted by a grey background on the MDR bar plot. For example, for the interaction (\texttt{rs1024216}, \texttt{rs1423304}), the genotype combinations (3,1), (2,2), (2,3), and (1,2) are associated with high-risk for PCOS. Higher order models exceed the limitations of succinct plotting and are omitted. 

\newpage

\section{Concluding Remarks}\label{discuss}
In this paper we have addressed the important issue of interaction screening in ultrahigh dimensional feature spaces. Although applications of interaction screening are wide spread, few extant methods exist for doing such. We have introduced a novel interaction screening method called JCIS (Joint Cumulant Interaction Screening) that is empirically accurate, theoretically sound, and computationally feasible.  

One unrivaled advantage of JCIS when compared to existing interaction screening methods such as iFORM (\cite{HaoZhangiFORM}) and GPC (\cite{Huang2014}) is the ability of JCIS to determine interactive effects among predictors even when no no strong marginal effects exist. 
Extant methods for feature interaction screening are deleteriously over-reliant on the existance of pronounced and explicit main effects from both an empirical and theoretical standpoint. The superiority of JCIS in this regard is born out repeatedly in the simulations of Section \ref{simulations}.  

Our proposed method also has the strong sure screening consistency property, meaning that even as the number of covariates increases exponentially with respect to sample size, JCIS prevails in discovering the exact set of relevant features with probability approaching one. This property has become the benchmark theoretical property for feature screening methods. The proofs pertaining to strong sure screening of JCIS are presented in Section \ref{proofSection} found below.  

Via a real data analysis on an empirical data set relating to polycystic ovary syndrome (PCOS) from the NCBI databases, we demonstrated the ability of our method to be applied to extremely large real life data sets such as those found in genetics. In terms of number of covariate pairs in question, as well as the number of observations considered, the real data set we  examine is (by conservative estimates) about 2800 times larger than the data sets examined in similar papers (e.g. the inbred mouse microarray gene expression dataset found in \cite{HaoZhangiFORM}). In turn, this means that the computational considerations necessary for our PCOS data were, until now, unseen in the setting of interaction feature screening. As both data dimension and computational power continue to grow, we feel confident that JCIS will remain a salient approach for analyzing two-way interactions in ultrahigh (and beyond) dimensional data sets.  

%%%\begin{comment}

\section{Proofs of Theoretical Results}\label{proofSection}

Here we present in full the proofs for Theorems \ref{Thm1} and \ref{Thm2}. Before proceeding into the proofs, we will establish a lemma which employs the Continuous Mapping Theorem (see \cite{MannWald1943} and \cite{Casella2002statistical}).
%%%%%%%%%%%%%%%%%%%%%%%%%%%%%%%%%%%%%%%%%%%%%%%%%%%%
%%%% FIRST LEMMA
%%%%%%%%%%%%%%%%%%%%%%%%%%%%%%%%%%%%%%%%%%%%%%%%%%%%

\subsection{Prefacing Lemmas and a Definition} The following lemmas will assist in the proof of our main theorems on strong sure screening.

\begin{lemma}\label{Lemma}
  Let $\hat{\sigma}_{j_1}$, $\hat{\sigma}_{j_2}$, and $\hat{\sigma}_Y$ be the estimators of $\sigma_{j_1}$, $\sigma_{j_2}$, and $\sigma_Y$ used in the definition of $\Rhat$.
Assume that $\hat{\sigma}_{j_1}$, $\hat{\sigma}_{j_2}$, $\hat{\sigma}_Y$, and $\hat{\tau}_{j_1, j_2}$ are all (individually speaking) \textit{consistent} estimators of the respective values they are estimating (viz. $\sigma_{j_1}$, $\sigma_{j_2}$, $\sigma_Y$, and $\kappa_3(X_{j_1}, X_{j_2}, Y))$.
We then have that \[\Rhat = \frac{\hat{\tau}_{j_1, j_2}}{\hat{\sigma}_{j_1}\hat{\sigma}_{j_2}\hat{\sigma}_Y}\] is a consistent estimator of $R_{j_1, j_2}$.
\begin{proof}
The proof of this lemma follows easily from a direct application of the Continuous Mapping Theorem. %paired with a straight forward generalization of the very similar proof found in \cite{mainEffects}. %%%%%%Generalization means more than just "categorical." %%%
\end{proof}
\end{lemma}
%%%\textit{[See \cite{Serfling1980}, Theorem of Section 1.7].}
%%Are all continuous functions on $\mathbb{R}^k$ Borel functions? Yes. This seems to be a %%%classical result.
\subsubsection{Classical results} It is a classical result that \[\hat{\sigma}_{j_1} = \frac{1}{n}\sum_{i = 1}^n \left(X_{ij_1} - \bar{X}_{j_1}\right)^2\] is a biased, yet consistent, estimator of the standard deviation of $X_{j_1}$. Similar statements can of course be made for $\hat{\sigma}_{j_2}$ and $\hat{\sigma}_Y$.

%%%%%%%%%%%%%%%%%%%%%%%%%%%%%%%%%%%%%%%%%%%%%%%%%%%%
%%%% PROOF OF MAIN THEOREMS
%%%%%%%%%%%%%%%%%%%%%%%%%%%%%%%%%%%%%%%%%%%%%%%%%%%%

\subsection{Proofs of Theorems \ref{Thm1} and \ref{Thm2}}

The proof of these two theorems is accomplished in three steps:
\begin{enumerate}
    \item We first show that a positive lower bound $R_{\min}$ exists for all $R_{j_1, j_2}$ with $(j_1, j_2) \in \mathcal{S}_T$. In other words, we will show the following:
     \[\text{There exists}~~ R_{\min} >0 \text{ such that } R_{j_1, j_2} > R_{\min}~~ \text{for all} ~~(j_1, j_2) \in \mathcal{S}_T.\]

    \item This is followed by our showing that $\Rhat$ is a uniformly consistent estimator of $R_{j_1, j_2}$ for each $1 \leq j_1 < j_2 \leq p$. This will effectually consist of showing that $\hat{\tau}_{j_1, j_2}$ is a consistent estimator of $\kappa_3(X_{j_1}, X_{j_2}, Y)$, since the standard deviation estimators in the denominator of $\Rhat$ are already well established consistent estimators of the standard deviations of $X_{j_1}$, $X_{j_2}$, and $Y$.

    \item We finally show that there exists a constant $c > 0$ such that
    \[\mathbb{P}(\widehat{\mathcal{S}} = \mathcal{S}_T) \longrightarrow 1 \text{ as } n \longrightarrow \infty\]
    Weak consistency is shown as a natural subcase of this, which will establish Theorem \ref{Thm2}.
\end{enumerate}

\subsubsection{Step 1}
We previously defined the following in Subsection \ref{thrtProp}:
\begin{eqnarray*}
    \omega_{j_1, j_2}(k_1, k_2, m) &=& (k_1 - \EE X_{j_1})(k_2 - \EE X_{j_2})(m - \EE Y)\pi_{j_1, j_2, Y}(k_1, k_2, m).
\end{eqnarray*}

\tR{Taking any fixed $X_{j_1}$ and $X_{j_2},$ let $f: \RR^3 \to \RR$ be defined by \[f(k_1, k_2, m) =\frac{\omega_{j_1, j_2}(k_1, k_2, m)}{\sigma_{j_1}\sigma_{j_2} \sigma_Y}.\] Since by Condition (C4) $R_{j_1, j_2}$ is finite for all $X_{j_1}$ and $X_{j_2}$ pairs, then this in turn implies that $f$ is integrable over $\RR^3$. We can thus state the following: 
\begin{align*}
  \infty > R_{j_1, j_2}  &= \int_{ \RR^3} f(k_1, k_2, m)~dk_1dk_2dm \\[1ex]
  & = \int_{\RR^3} \frac{\omega_{j_1, j_2}(k_1, k_2, m)}{\sigma_{j_1}\sigma_{j_2} \sigma_Y}~dk_1dk_2dm.
\end{align*}

%\newpage

Hence, for $(j_1, j_2) \in \mathcal{S}_T,$
\begin{eqnarray*}
R_{j_1, j_2} &=& \int_{\RR^3}\frac{ \omega_{j_1, j_2}(k_1, k_2, m)}{\sigma_{j_1}\sigma_{j_2} \sigma_Y}~dk_1dk_2dm \\[1.25ex]
&\geq& \frac{1}{\sigma_{\max}^3}\int_{\RR^3} \omega_{j_1, j_2}(k_1, k_2, m)~dk_1dk_2dm \quad \text{by (C2),}\\[1.25ex]
&\geq& \frac{1}{\sigma_{\max}^3} \sup\limits_{(k_1, k_2, m) \in \RR^3}\omega_{j_1, j_2}(k_1, k_2, m)\\[1.25ex]
&\geq& \frac{\omega_{\min}}{\sigma_{\max}^3} \quad \text{by (C3),}\\[1.25ex]
&>& 0.
\end{eqnarray*}
Define $R_{\min} = \dfrac{\omega_{\min}}{2\sigma_{\max}^3}.$
Then $R_{j_1, j_2} > R_{\min} > 0$ for all $(j_1, j_2) \in \mathcal{S_T}$.
\vspace{0.3cm}
This establishes a positive lower bound on $R_{j_1, j_2}$ for all  $(j_1, j_2) \in \mathcal{S_T}$, completing Step 1. Corollary \ref{Cor1} is also established by this step.}

\subsubsection{Step 2}\label{Step2}

 We now apply the weak law of large numbers to show that $\Rhat$ is a (uniformly) consistent estimator of $R_{j_1, j_2}$.
 This will consist of showing that $\hat{\tau}_{j_1, j_2}$ is a consistent estimator of $\kappa_3(X_{j_1},X_{j_2}, Y)$, since the denominator of $\Rhat$ is comprised of the routine (and, importantly here, consistent) estimators of $\sigma_{j_1}$, $\sigma_{j_2}$  and $\sigma_Y$.
 As it can be show using the Mann-Wald Theorem that the quotient of consistent estimators is itself a consistent estimator, our aforementioned work with $\hat{\tau}_{j_1, j_2}$ will suffice. This is reflected in the statement of Lemma \ref{Lemma}. 

By a slight rearrangement of the numerator in the definition of $\Rhat$, we can obtain
\begin{equation}\label{tauhat}
\hat{\tau}_{j_1, j_2}  =\frac{1}{n} \sum\limits_{i = 1}^n(X_{ij_1} - \overline{X}_{j_1})(X_{ij_2} - \overline{X}_{j_2})(Y_i - \overline{Y}).
\end{equation}

We now can explicitly expand the product of binomials in (\ref{tauhat}) to obtain
\begin{eqnarray*}
\hat{\tau}_{j_1, j_2} &=& \frac{1}{n}\sum X_{ij_1}X_{ij_2}Y_i - \frac{1}{n}\sum \overline{X}_{j_1}X_{ij_2}Y_i - \frac{1}{n}\sum X_{ij_1}\overline{X}_{j_2}Y_i \\\\
&&\quad\quad\quad\quad - \frac{1}{n}\sum X_{ij_1}X_{ij_2}\overline{Y} + \frac{1}{n}\sum \overline{X}_{j_1}\overline{X}_{j_2}Y_i\\\\
&&\quad\quad\quad\quad + \frac{1}{n}\sum \overline{X}_{j_1}{X}_{ij_2}\overline{Y}+ \frac{1}{n}\sum {X}_{ij_1}\overline{X}_{j_2}\overline{Y}-\frac{1}{n}\sum \overline{X}_{j_1}\overline{X}_{j_2}\overline{Y}
\end{eqnarray*}

By repeated applications (summand wise) of the weak law of large numbers to this above expression for $\hat{\tau}_{j_1, j_2},$ we obtain:
\[\frac{1}{n}\sum X_{ij_1}X_{ij_2}Y_i \xrightarrow{p} \mathbb{E}(X_{j_1}X_{j_2}Y)\]

\[\frac{1}{n}\sum \overline{X}_{j_1}X_{ij_2}Y_i \xrightarrow{p} \mathbb{E}(X_{j_1})\mathbb{E}(X_{j_2}Y)\]

\[\frac{1}{n}\sum {X}_{ij}\bar{Y} \xrightarrow{p} \mathbb{E}(X_j)\mathbb{E}(Y),\]
with all convergence being in probability. Similar conclusions can be reached for like terms. 

Hence we have
\begin{eqnarray*}
\hat{\tau}_{j_1, j_2} &=& \frac{1}{n}\sum X_{ij_1}X_{ij_2}Y_i - \frac{1}{n}\sum \overline{X}_{j_1}X_{ij_2}Y_i - \frac{1}{n}\sum X_{ij_1}\overline{X}_{j_2}Y_i \\\\
&&\quad\quad\quad - \frac{1}{n}\sum X_{ij_1}X_{ij_2}\overline{Y} + \frac{1}{n}\sum \overline{X}_{j_1}\overline{X}_{j_2}Y_i\\\\
&&\quad\quad\quad + \frac{1}{n}\sum \overline{X}_{j_1}{X}_{ij_2}\overline{Y}+ \frac{1}{n}\sum {X}_{ij_1}\overline{X}_{j_2}\overline{Y}-\frac{1}{n}\sum \overline{X}_{j_1}\overline{X}_{j_2}\overline{Y}\\\\
&\xrightarrow{p}& \EE(YX_{j_1}X_{j_2}) - \EE(YX_{j_1})\EE X_{j_2}-\EE(YX_{j_2})\EE X_{j_1}\\\\
&& \quad \quad \quad \quad \quad ~-\EE(X_{j_1}X_{j_2})\EE Y + 2 \EE Y\EE X_{j_1}\EE X_{j_2}\\\\
&=& \kappa_3(X_{j_1}, X_{j_2}, Y).
\end{eqnarray*}
So indeed $\hat{\tau}_{j_1, j_2}$ is a consistent estimator of $\kappa_3(X_{j_1}, X_{j_2}, Y)$. 
Furthermore, this shows, by Lemma \ref{Lemma}, that $\Rhat$ is a consistent estimator of $R_{j_1, j_2}$.

We will now show that such consistency is also uniform. 
Since $\Rhat$ is consistent as an estimator of $R_{j_1, j_2}$, we know that for any $1 \leq j_1 < j_2 \leq p$ and any  $\varepsilon > 0$,
\[\mathbb{P}(|\Rhat - R_{j_1, j_2}| > \varepsilon) \rightarrow 0 \quad \text{ as } n\rightarrow\infty.\]
Let \[(J_1, J_2) = \text{argmax}_{1 \leq j_1 < j_2 \leq p}~|\Rhat - R_{j_1, j_2}|.\]
Then, since $(J_1, J_2) \in \{1,2,\ldots, p\} \times \{1,2,\ldots, p\}$, we know that
\[\mathbb{P}(|\widehat{R}_{J_1, J_2} - R_{J_1, J_2}| > \varepsilon) \rightarrow 0 \quad \text{ as } n\rightarrow\infty\] for any $\varepsilon> 0$. In other words, we have that
\[\mathbb{P}\left(\max_{1 \leq j_1 < j_2 \leq p}|\Rhat - R_{j_1, j_2}| > \varepsilon\right) \rightarrow 0 \quad \text{ as } n\rightarrow\infty\] for any $\varepsilon> 0$.
This shows that $\Rhat$ is a \textit{uniformly} consistent estimator of $R_{j_1, j_2}$, completing Step 2. This also establishes Corollary \ref{Cor2}.

\subsubsection{Step 3}
%%%%(This follows \cite{Huang2014} closely).  %%%Keep this or not?

In Step 1 we defined $$R_{\min} = \dfrac{\omega_{\min}}{2\sigma_{\max}^3}.$$ Let $c = (2/3)R_{\min}$. Suppose by way of contradiction that this $c$ is insufficient to be able to claim $ \widehat{\mathcal{S}} \supseteq \mathcal{S}_T$. This would mean that there exists some pair $(j_1^*, j_2^*) \in \mathcal{S}_T$, yet $(j_1^*, j_2^*) \notin \widehat{\mathcal{S}}$. It then follows that we must have \[\widehat{R}_{j_1^*, j_2^*} \leq (2/3)R_{\min}\] while at the same time having (as shown in Step 1) \[R_{j_1^*, j_2^*}> R_{\min}.\]

From this we can conclude that \[|\widehat{R}_{j_1^*, j_2^*} - R_{j_1^*, j_2^*}| > (1/3)R_{\min},\] which implies that \[\max_{1 \leq j_1< j_2 \leq p}~|\Rhat - R_{j_1, j_2}| > (1/3)R_{\min}\] as well. However, we know by the uniform consistency of $\Rhat$ that by letting $\varepsilon = 1/3 R_{\min}$, we have
\[\mathbb{P}(\widehat{\mathcal{S}} \not\supseteq \mathcal{S}_T) \leq \mathbb{P}\left(\max_{1 \leq j_1 < j_2 \leq p}|\Rhat - R_{j_1, j_2}| >(1/3)R_{\min}\right) \rightarrow 0 \quad \text{ as } n\rightarrow\infty.\] This is a contradiction to the assumption of non containment above. So indeed, we have that \[\mathbb{P}(\widehat{\mathcal{S}} \supseteq \mathcal{S}_T)\rightarrow 1\quad \text{ as } n\rightarrow\infty.\]
This proves Theorem \ref{Thm2}, and also establishes the forward direction for the statement of Theorem \ref{Thm1}.

To prove the reverse direction for Theorem \ref{Thm1}, suppose (again by way of contradiction) that $\widehat{\mathcal{S}} \not\subseteq \mathcal{S}_T$. Then there is some $(j_1^*, j_2^*)\in \widehat{\mathcal{S}}$, yet $(j_1^*, j_2^*) \notin \mathcal{S}_T.$ This means that \[\widehat{R}_{j_1^*, j_2^*} \geq (2/3)R_{min},\] while at the same time (by (C4)) having \[R_{j_1^*, j_2^*} = 0.\] It now follows that \[|\widehat{R}_{j_1^*, j_2^*} - R_{j_1^*, j_2^*}| > (2/3)R_{\min}.\] Set $\varepsilon = (2/3)R_{\min}$. By uniform consistency we have
\[\mathbb{P}(\mathcal{S}_T \not\supseteq \widehat{\mathcal{S}}) \leq \mathbb{P}\left(\max_{1 \leq j_1< j_2 \leq p}|\Rhat - R_{j_1, j_2}| >(2/3)R_{\min}\right) \rightarrow 0 \quad \text{ as } n\rightarrow\infty.\]
From this we know that
\[\mathbb{P}(\mathcal{S}_T \supseteq \widehat{\mathcal{S}} )\rightarrow 1\quad \text{ as } n\rightarrow\infty.\]
We can now conclude that for $c = (2/3)R_{\min}$, we have $\mathbb{P}(\mathcal{S}_T = \widehat{\mathcal{S}}) \rightarrow 1$ as $ n\rightarrow \infty$, completing the proof.

%%%%\end{comment}

\bibliographystyle{imsart-nameyear}

\bibliography{BibliographyScreening}

\end{document}